\documentclass[traditabstract]{aa} 
\usepackage{graphicx}
\begin{document}
   \title{Relationship between wave processes in sunspots and quasi-periodic pulsations in active region flares}

   \author{R. Sych\inst{1}
        \and
        V.M. Nakariakov\inst{2}
	\and
	M. Karlicky\inst{3}
	\and
	S. Anfinogentov\inst{1}}
\institute{Institute of Solar-Terrestrial Physics,
           Irkutsk, Lermontov str., 126a, 664033, Russia\\
           \email{sych@iszf.irk.ru; anfinogentov@iszf.irk.ru}
	   \and
	   Physics Department, University of Warwick, Coventry, CV4 7AL, UK\\
           \email{V.Nakariakov@warwick.ac.uk}
     \and
           Astronomical Institute of the Academy of Sciences of the Czech Republic,
           25165 Onrejov, Czech Republic\\
           \email{karlicky@asu.cas.cz}}

\date{Received: {March 23}, 2009; accepted {August 06}, 2009}

\titlerunning{Sunspot oscillations and flaring QPP}
\authorrunning{Sych et al.}
\abstract
{A phenomenological relationship between oscillations in a sunspot and quasi-periodic pulsations in flaring energy releases at an active region above the sunspot, is established. The analysis of the microwave emission recorded by the Nobeyama Radioheliograph at 17~GHz shows a gradual increase in the power of the 3-min oscillation train in the sunspot associated with AR 10756 before flares in this active region. The flaring light curves are found to be bursty with a period of 3 min. Our analysis of the spatial distribution of the 3-min oscillation power implies that the oscillations follow from sunspots along coronal loops towards the flaring site. It is proposed that quasi-periodic pulsations in the flaring energy releases can be triggered by 3-min slow magnetoacoustic waves leaking from sunspots. }

\keywords{Waves -- Magnetohydrodynamics (MHD)}

\maketitle

\section{Introduction}
The problem of the transfer of energy, momentum, and information from sub-photospheric solar regions to both the corona and the solar wind is one of the most difficult in solar and stellar physics. Magnetohydrodynamic (MHD) waves are believed to play a key role because the waves are natural carriers of energy, momentum, and information (e.g., Erd{\'e}lyi 2006 for a recent review). In addition, the guided nature of the wave propagation opens up very interesting perspectives for fixing and tracing the energy transfer channels highlighted by the waves. In general, MHD waves can be guided by inhomogeneities in the characteristic MHD speeds (Alfv\'en, fast and slow) as well as the magnetic field itself, which are found in all regions of the solar atmosphere.

At the chromospheric level, one of the most pronounced wave phenomena are the 3-min oscillations over sunspots (see, e.g. Bogdan \& Judge 2006 for a recent review), usually detected as intensity oscillations in visible light, UV, and EUV spectral lines, as well as in microwave band (e.g., Shibasaki 2001, Gelfreikh et al. 1999, Nindos et al. 2002) and in the dm-radio flux records (M\'esz\'arosov\'a et al. 2006). Three-minute oscillations in sunspots are believed to be associated with slow magnetoacoustic waves (e.g., Zhugzhda 2008). Outwardly propagating compressible waves of the same periodicity are also seen in both the EUV 171\AA\ and 195\AA\ bandpasses in the magnetic fan structures situated over sunspots (e.g., De Moortel 2006 for a review). The projected phase speed of these waves is subsonic, and the waves are seen to propagate along the plasma channels elongated along the coronal magnetic field lines, and hence are interpreted as slow magnetoacoustic waves. Compressible 3-min waves observed at the same location in both EUV bandpasses show a high degree of correlation (King et al. 2003). The relationship between these waves and 3-min oscillations in sunspots remains unclear. The understanding of the propagation of 3-min oscillations through the solar atmosphere is one of the most important problems of solar physics. Its understanding will probably indicate the nature and properties of the plasma channels that transfer these waves into the corona, and hence the connectivity of different layers of the atmosphere. The role played by 3-min oscillations in the corona is also of interest, in particular, the relationship between 3-min oscillations in sunspots and the flaring activity in the active regions (AR) above the sunspots. A possible indication of such a relationship was mentioned in Gelfreikh (2002).

Wave and oscillatory phenomena in various parts of the solar atmosphere can trigger and modulate bursty energy releases, e.g., solar flares. In this case, the periodicity of the oscillations will be evident in the flaring light curves as quasi-periodic pulsations (QPP). This can be achieved by several mechanisms.
In the scenario proposed by Nakariakov et al. (2006), energy of transverse (kink or sausage) oscillations of coronal loops can periodically leak to a magnetic neutral point or line situated nearby. The incoming fast magnetoacoustic wave refracts towards the neutral point, experiencing focussing and steepening. This periodically generates very sharp spikes of electric current density in the vicinity of the neutral point, which in turn can be affected by current driven plasma micro-instabilities. The instabilities can cause the onset of micro-turbulence and hence enhance the plasma resistivity by several orders of magnitude. This would lead to periodic triggering of magnetic reconnection and hence the manifestation of the loop oscillations as periodic variation in the flaring light curve.

A compressible wave can periodically trigger magnetic reconnection not only by periodic current density spikes, but also by the variation in the plasma density in the vicinity of the reconnection site. This possibility was modelled numerically by \cite{2006SoPh..238..313C} in interpreting 3--5~min periodicity detected in repetitive bursts of explosive events in the transition region (Ning et al. 2004). Density variations result in a periodic variation in the electron drift speed. Depending upon the ratio of electron to proton temperatures, the value of the speed controls the onset of the Buneman or ion-acoustic instabilities and hence anomalous resistivity. The periodic onset of the anomalous resistivity triggers periodic energy releases. Transverse compressible waves may also directly trigger magnetic reconnection causing transition region explosive events, by changing the magnetic field strength (Doyle et al. 2006). Longitudinal, e.g., acoustic waves can also modulate flaring energy releases, either directly by the modulation of the drift velocity or the modulation of gyrosynchrotron emission efficiency (Nakariakov \& Melnikov 2006), or indirectly e.g., by means of centrifugal conversion into fast magnetoacoustic waves on the curved magnetic field lines (Zaitsev \& Stepanov 1989).

Kislyakov et al. (2006) analysed 15 flares observed in the 37~GHz band with the Mets\"ahovi radio telescope (Finland) with the use of the \lq\lq sliding window" Fourier transform and the Wigner--Ville nonlinear transform. The telescope spatial resolution is 2.4~arc min, the sensitivity is about 0.1 sfu, and the time resolution was higher than 0.1~s. During 13 events (about 90\%), a 5-min periodic modulation of the emission intensity was detected with the frequency of $3.2\pm 0.37$~mHz. In addition, a shorter period (about 1~s) signal was detected, which was found to be frequency modulated with the same 5-min period. In the development of this study (Za{\u i}tsev \& Kislyakov 2006), simultaneous modulation of the microwave emission by three low frequency signals with periods of 3.3, 5, and 10~min was observed in 30\% of the analysed outbursts. It was suggested that the detected modulation was caused by the parametric resonance between 5-min velocity oscillations in the solar photosphere and natural acoustic oscillations of coronal magnetic loops modulating the microwave emission. The detected periods of 5, 10, and 3 min were interpreted to correspond to the pumping frequency, its subharmonic, and its first upper frequency of parametric resonance, respectively. Confirmation of this finding with a different instrument is required. The physical mechanisms responsible for the appearance of 3 and 5 min flaring QPP cannot be determined without spatial information.

In general, spatial information, such as the spatial size and shape of the region occupied by an oscillation, and the distribution of the oscillation power, phase, and spectrum over the source, is crucial to establishing the nature of the QPP (e.g., Grechnev et al. 2003; Melnikov et al. 2005). Novel imaging data analysis techniques developed in solar physics have been shown to allow one to exploit the full potential of the spatially resolving observations. Grechnev (2003) proposed creating a 2D broadband variance map, representing the overall dynamics of an analysed event, from microwave correlation data cubes obtained with NoRH. This approach allows us to ascertain the spatial locations of faint variable microwave emission sources.
Nakariakov \& King (2007) designed a coronal-periodmapping technique, which reduces 3D imaging data cubes (2D in space and time) to a sequence of static maps inferring collective oscillations of extended (larger than the pixel) coronal structures. This approach was successfully applied by Inglis et al. (2008) to the study of single-periodic pulsations in a large off-limb flaring loop seen in the microwaves with NoRH. However, period-mapping does not provide any phase information, and does not allow detailed studying of multi-periodic or non-stationary phenomena. Complex temporal and spatial features of oscillatory processes in imaging datasets can be studied with the pixelised wavelet filtering (PWF) technique (Sych \& Nakariakov 2008). This approach produces 4D (2D in space, time and frequency) data cubes providing information about the time modulation of oscillatory signals, their coupling, and their evolution. In particular, this method allows one to obtain information about the spatial structure of narrowband and broadband time signals, as well as the analysis of the signal integrated over the whole field-of-interest. The practical implementation of PWF consists of several steps:
 \begin{itemize}
   \item The images in the time sequence are coaligned, removing the spatial mismatch between consecutive images.
   \item The object of investigation (e.g., the period of the oscillations) and the field-of-interest (FOI) are selected.
   \item Construction of the variance map of the FOI, determining the spatial distribution of the integrated power of the time signal.
   \item Direct wavelet transform in the time domain, filtering out the spectral components of interest, and inverse wavelet transform. The resultant data cube contains the time variation in the signal in the prescribed spectral band.
 \end{itemize}
In addition, global wavelet spectra can be calculated for each pixel, resembling the construction of a periodmap. PWF is a convenient tool for establishing a relationship between different spatially-separated oscillatory processes.

Analysis of the possible relationship between flaring energy releases and dynamical processes in the lower regions of the solar atmosphere can shed light on the triggering of the energy releases and hence the basic physical processes responsible for them. The spatially-resolved analysis of this relationship can also infer the atmospheric connectivity channels. The aim of the paper is to establish a phenomenological relationship between dynamical processes occurring in a sunspot and in flares, within the active region (AR) linked to the sunspot, and identify the channels of the connectivity. We demonstrate that 3-min oscillations of a sunspot appear to be present in the microwave emission associated with the flaring activity over the sunspot.

\section{Observations}
\label{sec:obs}

\begin{figure}
\resizebox{\hsize}{!}{\includegraphics[viewport=68 370 463 729]{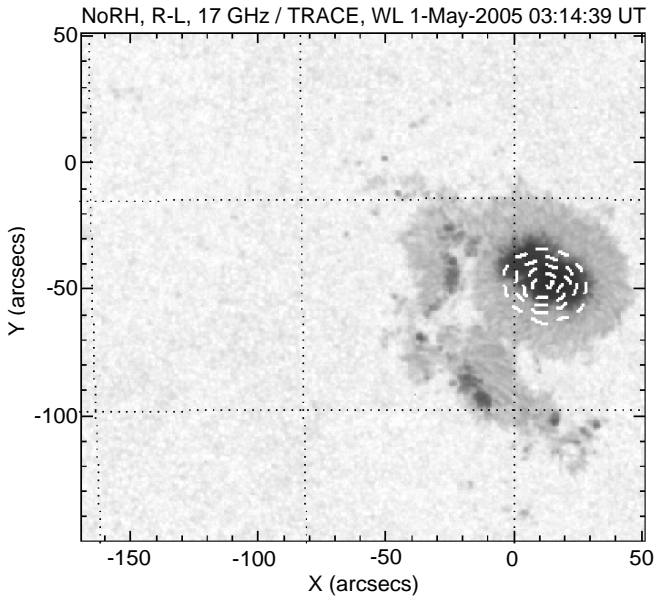}}
\caption{Combined image of AR 10756 taken with NoRH at 17 GHz and in the optical band with TRACE on 2005 May 01 at 03:14:39~UT. The spatial distribution of the microwave circular polarisation (contours) overlaps with the TRACE white light image (background).}
\label{combim}
\end{figure}

We investigate 3-min oscillations in the NOAA sunspot group 10756 during its passage through the solar disk from 2005 April 28 to 2005 May 4, and the apparent manifestation of these oscillations as QPP of the flaring microwave emission generated in this AR. During all observations, the AR was the only one on the solar disk. The AR consisted of a large symmetric leading spot of negative polarity and a trailing positive polarity plage with several small size spots and pores (see Fig.~1 and also Fig.~2 in Denker et al. 2007). During the disk passage, the trailing part reduced in size and gradually disappeared. There is a strong L-polarised microwave source observed with the Nobeyama Radioheliograph (NoRH) in the 17~GHz channel during the complete time interval of the disk passage. The circular polarisation (V=R-L) source is situated over the sunspot umbra (see Fig.~\ref{combim}). The flaring activity of this AR is manifested mainly by weak bursts of duration from 5 to 30 min, and also with the appearance of short, 3-10 s, pulses or spikes. The majority of the spike-like events happened on the 1st of May, 2005, when the AR was situated at the central meridian. A possible reason for this feature is considered in our discussion.

\begin{figure}
\resizebox{\hsize}{!}{\includegraphics[viewport=0 170 535 789]{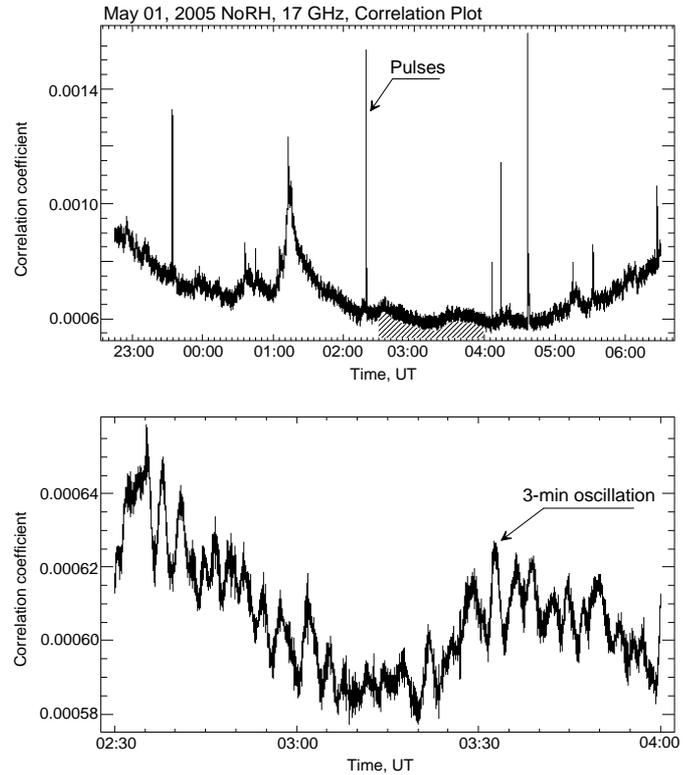}}
\caption{An example of the flaring activity on 2005 May 1: the correlation time profile of NoRH at 17~GHz. The bottom panel zooms the time interval hatched in the upper panel. The arrows indicate the dynamical features of the signal discussed in the paper.}
\label{lightcurve0105}
\end{figure}

\begin{figure}
\resizebox{\hsize}{!}{\includegraphics[viewport=76 47 480 787]{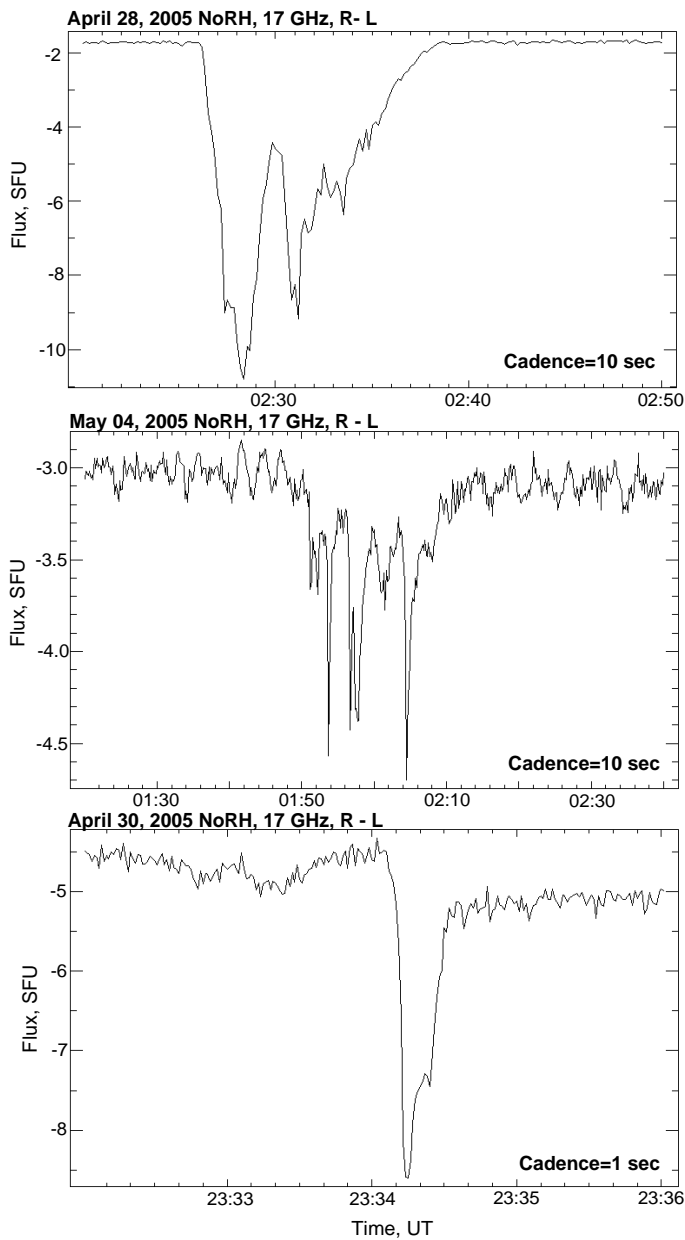}}
\caption{The microwave burst fluxes profiles of the active region 10756 at 17~GHz recorded with NoRH in the circular polarisation channel (R-L) on 2005 April 28 (upper panel), 2005 May 4 (central panel), and 2005 April 30 (bottom panel). The events shown in the two upper panels are considered as examples of microwave bursts, while the event in the lower panel is an example of a microwave spike-like pulse.}
\label{burst_lc}
\end{figure}

Figure~\ref{lightcurve0105} shows an example of a typical daily full-disk microwave correlation time profile obtained with NoRH. The correlation time profiles (TCX files) are connected with the spatial Fourier components of the microwave brightness distribution over the solar disk.
The definition of the correlation is \lq\lq averaged values of the correlation between antenna pairs (after removing short base-line pairs) of NoRH". The correlation increases as a strong localised microwave signal appears, produced by e.g., a flare. Visual inspection of the correlation profile shows the presence of bursts of various duration (at 00:35, 01:15, and 05:15 UT), and spike-like pulses that are solitary powerful peaks of intensity of short, less than 10 s, duration (at 23:35, 00:35, 00:45, 02:20, 04:07, 04:15, 04:37, and 05:33 UT). By zooming into a part of the smoothed correlation plot (the bottom panel of Fig.~\ref{lightcurve0105}), we can detect 3-min oscillations in the emission. Since the correlation signal is dominated by the change in the brightness of sources of small angular size, one can conclude that the source of the 3-min oscillations is the sunspot.

Figure~\ref{burst_lc} shows polarisation time profiles of the microwave bursts on 2005 April 28 (02:25-02:40 UT) and 2005 May 4 (01:50-02:10 UT) and a spike-like pulse on 2005 April 30 (23:34:15 UT). Visual analysis of the bursts infers that a quasi-periodic component is present with a period of about 3 min in both the pre-flare and burst stages. The apparent presence of this oscillatory component both before and during the energy release, motivated this study. The spike-like pulse was considered in the context of the investigation of the spatial localisation of its source and a possible relation with the sources of more long-living energy releases. In the following, we consider all three events in detail.

\subsection{The flare on 2005 April 28}
\label{2804}

\begin{figure}
\resizebox{\hsize}{!}{\includegraphics[viewport=0 280 518 800]{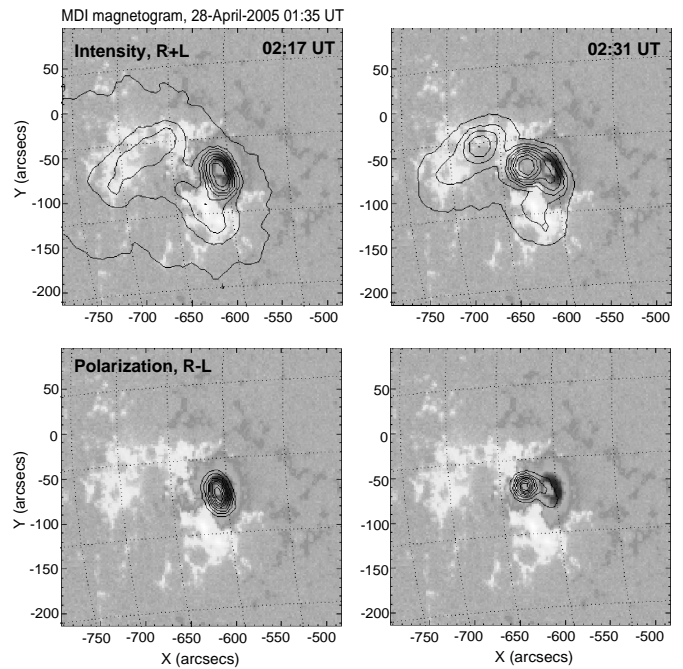}}
\caption{
The MDI magnetogram (01:35 UT, 2005 April 28) of AR 10756. The contours show the 17 GHz microwave sources obtained with NoRH in intensity and polarisation channels before the flare at 02:17 UT (left panels) and during the flare at 02:31 UT (right panels). The positive and negative magnetic polarities are shown by white and black colors, respectively.}
\label{comb280405}
\end{figure}

For the analysis of this flare, we used the 17~GHz microwave maps synthesised for two time intervals, before the flare (02:17~UT) and during the flare (02:31~UT) (see Fig.~\ref{comb280405}). The relation between the localisation of the magnetic field and the microwave sources is investigated by overlapping the microwave images with an MDI magnetogram of the AR, constructed before the flare at 01:35 UT. Before the flare, the microwave intensity source consists of two parts that spatially coincide with the location of the leading and trailing parts of the sunspot group. The leading part (-620,-70) is situated at the sunspot that has a strong field (about 2000 G) of negative polarity. The trailing part (-700, -50) coincides with the positive polarity plage region that has a field of about 800 G. During the flare, two microwave intensity sources appear with maxima situated over the local neutral lines in the leading (-640,-60) and trailing (-690,-40) parts of the AR. In the polarisation channel, there is a single highly polarised source close to the sunspot, similar to one shown in Fig.~\ref{combim}.

\begin{figure}
\resizebox{\hsize}{!}{\includegraphics[viewport=69 383 423 738 ]{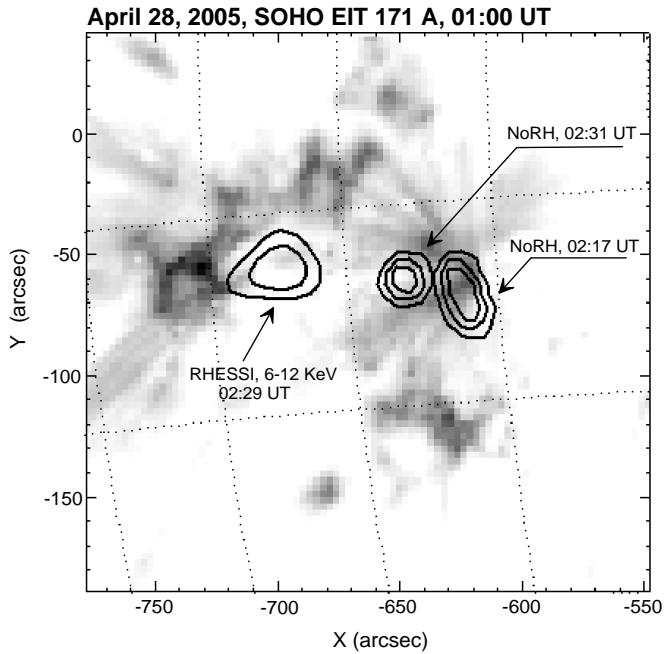}}
\caption{The EUV coronal loop structure of the AR 10756 on 2005 April 28 at 01:00 UT obtained with SOHO EIT (171\AA\ ), and locations of the NoRH polarisation (R-L) microwave source before (02:17 UT) and during (02:31 UT) the flare.  The X-ray (RHESSI, 6--12 KeV) source at 02:29~UT is shown by contours. Arrows indicate the spatial locations of the sources.}
\label{comb280405_2}
\end{figure}

The EUV image of this AR obtained with SOHO/EIT in the 171\AA\ bandpass (Fig.~\ref{comb280405_2}) about eighty minutes before the flare, shows the presence of a system of loops connecting the leading sunspot to the trailing plage (c.f. the MDI image in Fig.~\ref{comb280405}).
The X-ray image of this active region obtained with RHESSI in the 6-12 keV energy channel during the flare (at 02:29 UT), shows the presence of a bright X-ray source  that coincides with the fainter microwave intensity source over the plage (see Fig.~\ref{comb280405}).

The observed mismatch of the spatial location of microwave and X-ray sources can be connected with the directivity of the microwave emission (Bastian et al. 1998; Nindos et al. 2008). The observed structure of the spatial location of microwave 17 GHz and X-ray sources can also be explained by emission from an asymmetric magnetic trap (the magnetic field in one footpoint differs from the second one). While the gyro-synchrotron emission of superthermal electrons is preferentially generated at the footpoint with a higher magnetic field, the X-ray emission is dominant on the opposite side because of the easier precipitation of these electrons into the denser chromospheric layers.

Temporal and spatial features of the oscillatory processes in the AR are studied in more detail with the use of the PWF method with the Morlet mother function, applied to time sequences of 2D images of the emission polarisation (R-L) synthesised from the 17 GHz NoRH data in the time interval 01:30-03:00 UT. The time resolution was 10 s and the spatial resolution was 10". The FOI size was 75"$\times$75" and included the image of the sunspot.

\begin{figure}
\resizebox{\hsize}{!}{\includegraphics[viewport=0 506 524 782]{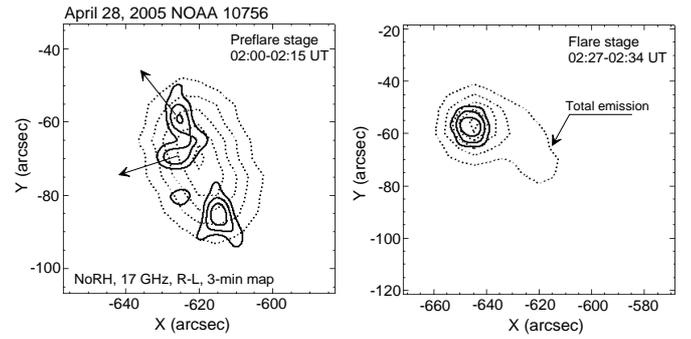}}
\caption{ The solid contours show the spatial structure of the narrowband 3-min oscillation microwave sources for different phases of the flare on 2005 April 28: in the pre-flare phase (02:00-02:15 UT; left panel) and during the flare (02:27-02:34 UT; right panel). The dotted contours show the total (broadband) microwave emission. The arrows shows the possible paths of the running waves from the sunspot.}
\label{tp280405}
\end{figure}

\begin{figure*}
\resizebox{\hsize}{!}{\includegraphics[viewport=0 426 553 766]{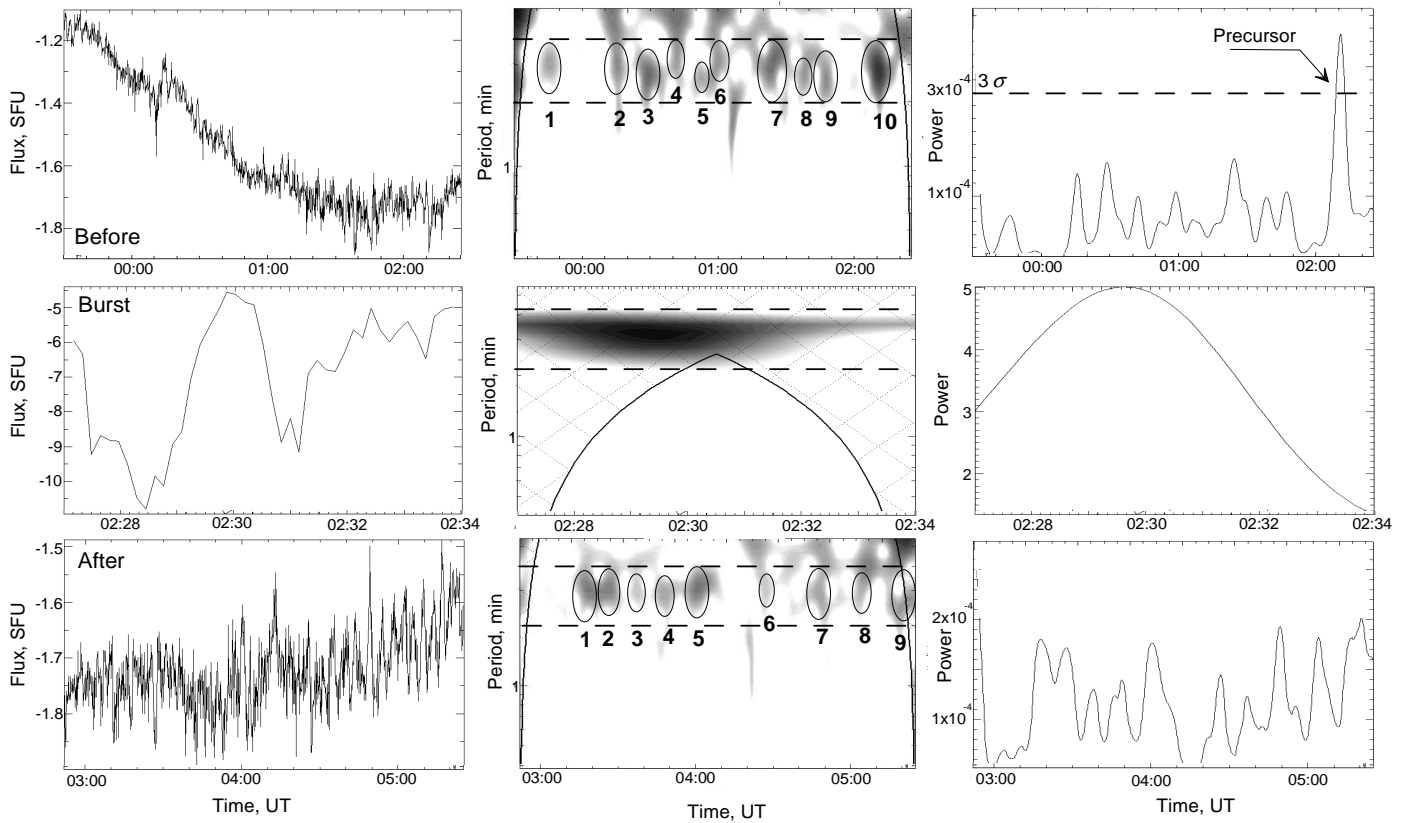}}
\caption{Time profiles of the integrated polarisation signal (left column), their power wavelet spectra (central column), and time profiles of the 3-min component (right panel) of the event on 2005 April 28. In the wavelet spectra, the spectral range corresponding to the periods from 2 min to 4 min is shown by the horizontal dashed lines. The hatched region exhibits the cone-of-influence. In the right panel, the horizontal dashed line indicates the confidence level of 3$\sigma$. The upper row of panels corresponds to the time interval before the flare (23:30-02:25 UT), the middle row shows the signal during the flare (02:27-02:34 UT), and the bottom panel shows the signal of the decay phase (02:35-05:25 UT). The time is given in min. }
\label{fig7}
\end{figure*}

We now consider the spatial structure of the 3-min oscillations and its evolution during the event. Using the PWF technique, we construct maps of the narrowband 3-min oscillations for two different phases of the event: during the significant increase in the oscillation power before the flare, and during the flare (see Fig.~\ref{tp280405}), as shown in Fig.~\ref{fig7}. It is evident that during the growth of the oscillation power before the flare (Fig.~\ref{fig7}, left panel), the narrowband source consists of two parts situated above and below (in the figure) the centre of the broadband emission. The latter is located above the umbra-penumbra boundary. Both upper and lower parts have an asymmetric V-shape. The size of these parts is close to the threshold of the instrument resolution (about 10"), although, the use of dynamical information allows us to localise the source (see also the discussion in Inglis et al. 2008). The arms of the V-shaped sources seem to be parallel to the loop system seen in EUV (Fig.~\ref{comb280405_2}). This most likely indicates the leakage of 3-min oscillations from the sunspot in the direction of the trailing part of the AR along the coronal magnetic fan structures.

The right panel of Fig.~\ref{tp280405} shows the spatial source of the 3-min oscillations during the flare. It has a symmetric form, and is situated above the ends of the upper V-shaped source found in the pre-flare phase. This spatial coincidence of the 3-min oscillation sources determined before and during the flare confirms their apparent relationship. Thus, the results obtained can be interpreted as an indication that the periodicity observed in the flaring energy releases is somehow generated in the sunspot.

Our analysis of the polarisation signal integrated over the FOI for the time intervals before the flare (23:40-02:25 UT), during the flare (02:27-02:34 UT), and during the decay phase (02:50-05:25 UT) is presented in Fig.~\ref{fig7}. The figure shows the time profiles (light curves) of the integrated flux, the wavelet spectra of these signals (the spectral range corresponding to periods from 2 min to 4 min is shown by the horizontal dashed lines), and time profiles of the power of the 2-4 min spectral components. According to the upper panels of Fig.~\ref{fig7}, before the flare the 3-min oscillations were grouped into wave trains of typical duration about 10-12 min. In the figure, the ellipses indicate the times of the oscillation trains, which are also labeled by consecutive numbers. In the pre-flare phase, the amplitude of the trains grows gradually, reaching its maximum at about 15 min before the flare. This is highlighted in the top right panel of Fig.~\ref{fig7} which shows the time profile of the narrowband spectral component corresponding to periods from 2 min to 4 min. The arrow indicates the time localisation of the burst precursor.

Before the flare, the power of 3-min oscillations exceeds the level of 3$\sigma$. The observed phenomenology could be interpreted as evidence of some association between the 3-min oscillations and the flare. In particular, it could represent the triggering of the flare by the 3-min oscillation power guided by the magnetic field from sub-photospheric layers, when the oscillation has sufficiently high amplitude. A similar behaviour was found for the other flare, which occurred on 2005 May 5, and is discussed below.

The middle panels of Fig.~\ref{fig7} show that the flaring phase is characterised by three consecutive bursts also of period about 3 min. The power of the 3-min oscillations during the flare is two orders of magnitude higher than in the pre-flare phase. In the decay phase shown in the bottom panels of Fig.~\ref{fig7}, the 3-min oscillations are also present. Their level is of the same order as in the pre-flare phase, and does not exceed the 3$\sigma$ level.

\subsection{The flare on 2005 May 4}

\begin{figure*}
\resizebox{\hsize}{!}{\includegraphics[viewport=22 445 522 756]{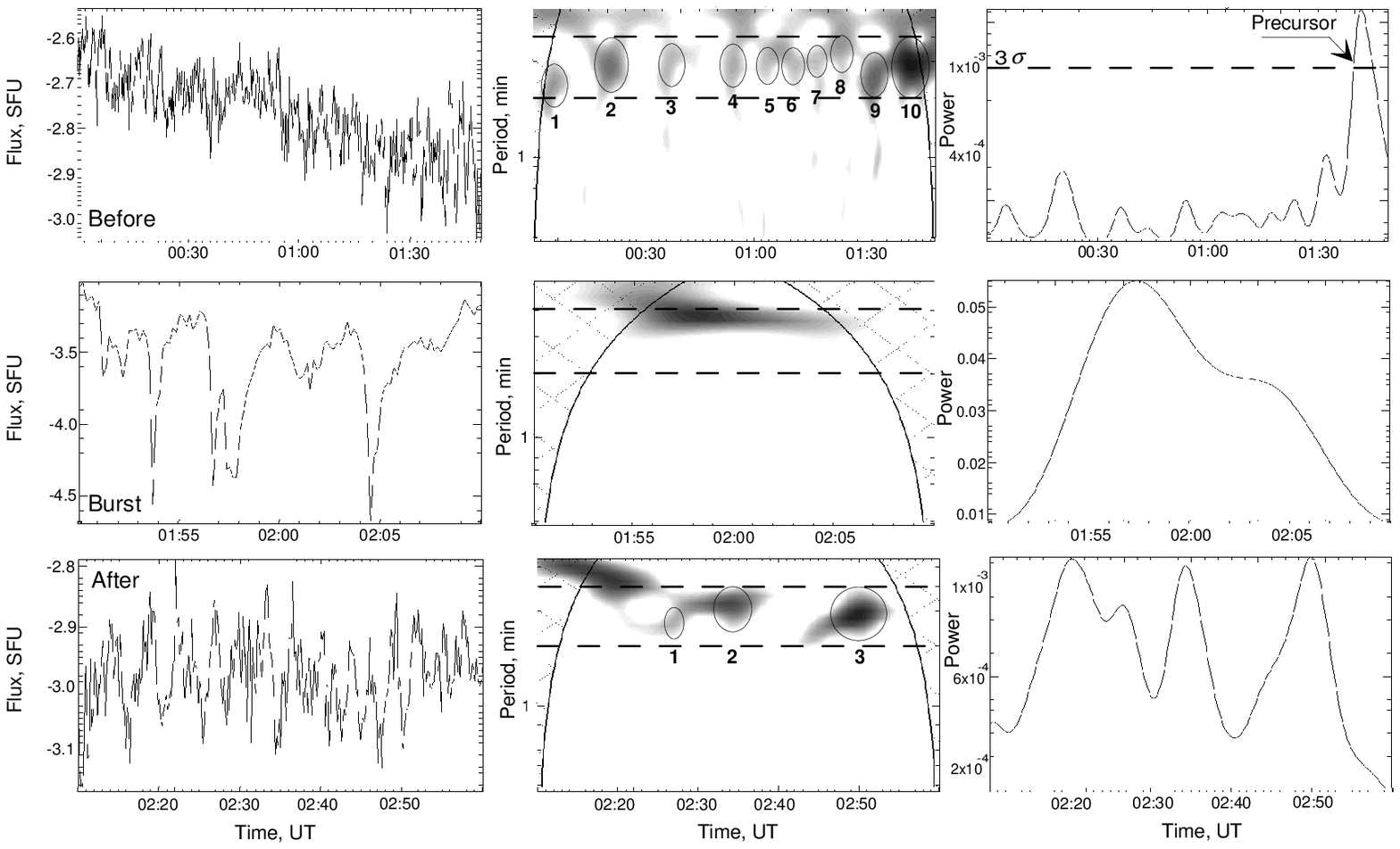}}
\caption{Time profiles of the integrated polarisation signal (left column), their power wavelet spectra (central column), and time profiles of the 3-min component (right panel)  of the event on 2005 May 4. The upper row of panels corresponds to the time interval before the flare (00:00-01:50~UT), the middle row shows the signal during the flare (01:50-02:10~UT), and the bottom panel shows the signal of the decay phase (02:10-3:00~UT).}
\label{fig8}
\end{figure*}

We consider another flaring event associated with AR~10756, which occurred on 2005 May 4 at 01:50-02:10~UT. The visual inspection of the light curve of this event (Fig.~\ref{burst_lc}, central panel) implies that 3-min oscillations are present. As in the event on 2005 April 28, we first analyse the time evolution of the polarisation signal integrated over the FOI. The results obtained for the integrated signal are presented in Fig.~\ref{fig8} in a similar fashion to Fig.~\ref{fig7}. As on 2005 April 28, 3-min oscillation trains are present in all phases of the flare. In the wavelet spectrum, the filling period of these wave trains is situated in the 2-4 min band and is not seen to change with time. Before the flare, in the time interval 01:36 - 01:50 UT one can see a strong, well over $3\sigma$, increase in the power of 3-min oscillation trains, with a maximum at about 01:40~UT (see Fig.~\ref{fig8}, upper row). During the flaring phase (01:50-02:10~UT), there were four pronounced emission peaks of period about 3 min (see  Fig.~\ref{fig8}, middle row).  The maximum power of 3-min oscillations was reached at about 1:57~UT.
In the post-flare phase (02:10-03:00~UT, see  Fig.~\ref{fig8}, bottom row), when the emission decreases gradually to the pre-flaring level, trains of 3-min oscillations are also present. At this stage, the 3-min oscillation trains have approximately constant power. The time intervals between the oscillation trains are in this case about 10-20 min. We emphasize that this behaviour appears to differ from the post-flare phase of the event on 2005 April 28 (Sect.~\ref{2804}), when initially the time intervals between the 3-min oscillation trains were rather stable with a value of about 12-13 min.

\begin{figure}
\resizebox{\hsize}{!}{\includegraphics[viewport=48 347 467 740]{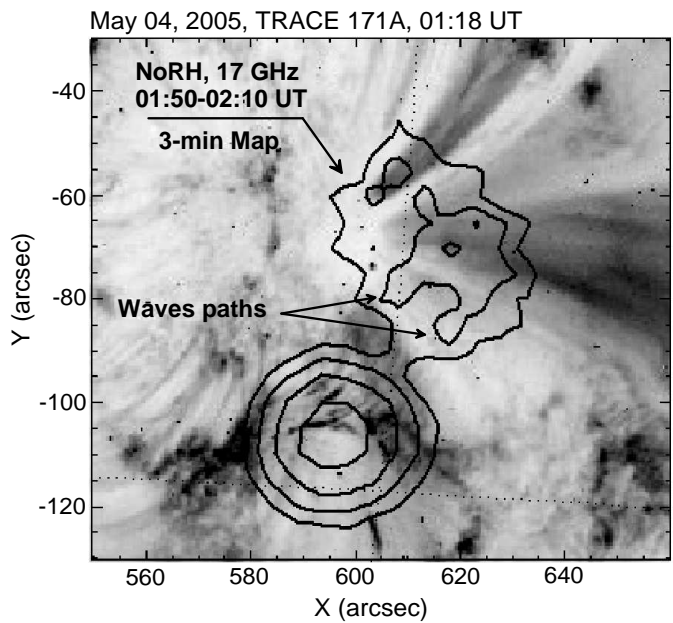}}
\caption{The EUV coronal loop structure of the AR 10756 on 2005 May 4 at 01:18~UT obtained with TRACE in 171\AA\ . The V-like contours show the locations of the sunspot narrowband (3~min) microwave polarisation source during the burst at 01:50-02:10~UT. The spherical source corresponds to the burst. The arrows indicate the possible paths of the waves.}
\label{fig9}
\end{figure}

Figure~\ref{fig9} shows the spatial location of microwave sources during the flare. There are two distinct sources, one situated over the sunspot and another associated with the flare. This time, in contrast to the event on 2005 April 28, the sunspot and the flare-associated burst sources of the microwave emission are spatially separated at a distance of about 35" from each other. Hence, this event allows us to study 3-min oscillations over the sunspot and in the flare site, spatially separately but simultaneous in time.

As in the previous section, the spatial structure of 3-min oscillations is analysed by the PWF method. Figure~\ref{fig9} shows the location of the narrowband 3-min oscillation maps superimposed on the EUV image of the active region, which highlights the structure of the coronal magnetic field. Both the sunspot atmosphere and the flaring energy release show pronounced 3-min oscillations.  It is seen that while the flare source is symmetric, the sources of 3-min oscillations over the sunspot have a pronounced V-shape. The arms of the V-shaped structure are more pronounced during the maxima of the 3-min oscillation trains. This is similar to the event discussed in the previous section, but the arms of the V-shaped structure spread in a different direction. In contrast to the event on 2005 April 28, the arms are seen to extend in the direction perpendicular to the apparent plasma structures in the magnetic fan extended from the sunspot. However, this interpretation is counterintuitive, because 3-min waves radiating from a sunspot into the corona are known to be guided by magnetic field lines (e.g., De Moortel, 2006). Another option could be fast waves, but their localisation in the direction perpendicular to their wave fronts requires the presence of some guiding structures (e.g., Van Doorsselaere et al., 2008). On the other hand, the direction of the arms coincides with the direction of loops in the EUV arcade that spatially coincide with the flare source in the image (see Fig.~\ref{fig9}). Since the EUV loops should be situated under the flare site, it is reasonable to assume that the arms of the V-shaped structure of narrowband 3-min oscillations highlight hotter magnetic structures (which are not seen in EUV) situated over the EUV arcade, which may link the flare site to the sunspot. Hence, as for the event on 2005 April 28, the arms may represent the paths of 3-min wave leakage from the sunspot before the flare, along the magnetic field lines towards the flare epicentre.

\begin{figure}
\resizebox{\hsize}{!}{\includegraphics[viewport=15 250 531 780]{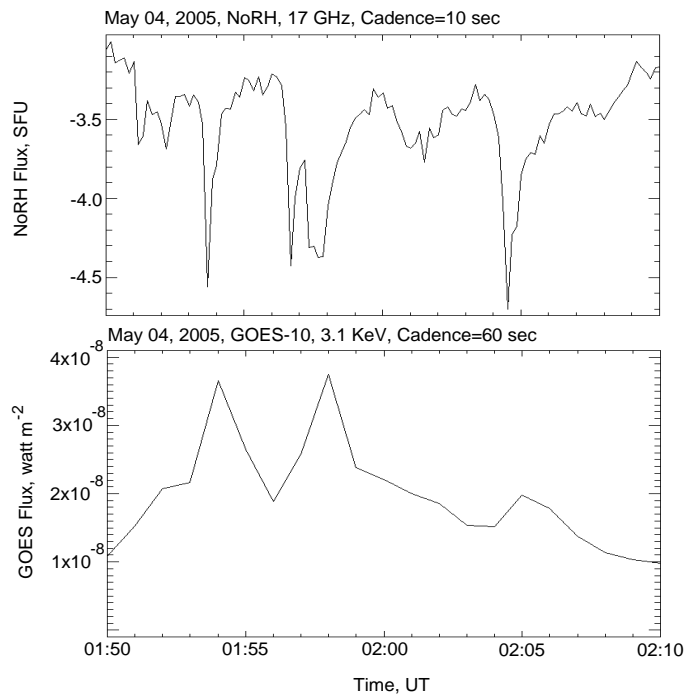}}
\caption{The comparison of the microwave (NoRH, R-L, 17 GHz, 10 s cadence) and soft X-ray (GOES-10, 3.1 KeV, 60 s cadence) light curves of the solar flare on 2005 May 4.}
\label{nob_com10}
\end{figure}

In this case, the fingers of the V-shaped structure provide seismological evidence of the waveguiding structure unseen in EUV. Unfortunately, no imaging information corresponding to hotter temperature was available for this event. In particular, the RHESSI spacecraft was in the radiation belts (01:45-02:19 UT).  However, the soft X-ray light curve recorded by GOES-10 in the 3.1 keV channel with 60 s cadence show the presence of 3 min modulation. Figure~\ref{nob_com10} compares the 17 GHz microwave light curve obtained in the polarisation channel (10 s cadence) with the GOES-10 soft X-ray emission curve. The curves are clearly anticorrelated. The absolute value of the correlation coefficient, calculated after rebinning the microwave signal to the 60 s cadence, was found to be 0.7. Thus, we can deduce that the 3 min waves observed to propagate along the V-shaped structure, produce the 3-min modulation of the soft X-ray emission from the same, (invisible in EUV), structure. However, direct confirmation of this structure can be achieved only by imaging data.

\begin{figure}
\resizebox{\hsize}{!}{\includegraphics[viewport=4 257 505 513]{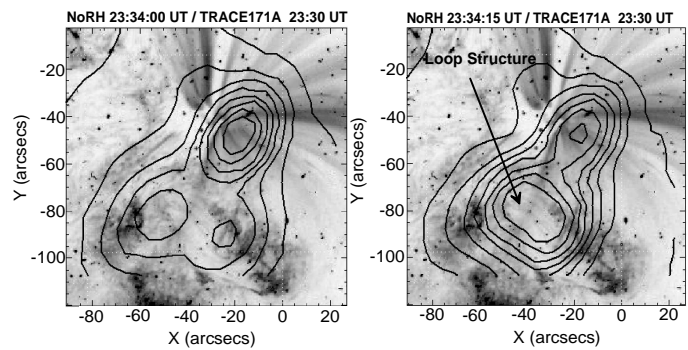}}
\caption{Microwave polarisation images (NoRH, 17 GHz, shown by contours) taken at the beginning (23:34:00 UT) and at the maximum (23:34:15 UT) of the spike-like pulse on 2005 April 30, superimposed on the EUV image obtained with TRACE at 171\AA\ at 23:30 UT. The arrow indicates the location of the coronal loops near the magnetic neutral line.}
\label{fig11}
\end{figure}

\subsection{The spike-like pulse on 2005 April 30}

As it has been pointed out, the flaring activity on 2005 April 28 -- 2005 May 04 is also characterised by the occurrence of short, of duration of about 10-30~s, spike-like pulses of microwave emission (see e.g., Fig.~\ref{burst_lc}). Here, we study a single spike-like pulse that occurred on 2005 April 30 at 23:34~UT, and its possible relationship with the wave processes in the sunspot. The spike-like pulse, shown in the bottom panel of Fig.~\ref{burst_lc}, has a rather simple time signature consisting of a sharp increase in the microwave polarisation emission and then a quick return of the polarisation signal to approximately the pre-pulse level. The total duration of this pulse is about 25~s.
This behaviour is typical of microwave spikes (which are usually of far shorter duration, less than 1~s; see e.g., Altyntsev et al. 1996) and presumably associated with fast impulsive magnetic reconnection in a complex magnetic configuration.
Figure~\ref{fig11} shows the spatial structure of the microwave burst superimposed on the EUV image of the coronal plasma configuration. At the beginning of the pulse (Fig.~\ref{fig11}, left panel), the microwave emission comes from three distinct sources: the stronger one over the sunspot, and two weaker sources situated at about 30" from the sunspot, which appear during the spike-like pulse of microwave emission. The sunspot source is symmetric and slightly extended towards the pulse source. During the development of the pulse, its emission is seen to come from the region situated between the two initial sources (see Fig.~\ref{fig11}, right panel), which is also typical of microwave spikes (Altyntsev et al. 1996). This may indicate the formation of a loop with footpoints situated at the two initial microwave sources.
According to Fig.~\ref{fig11}, the spatial location of one of the footpoints is close to the epicentre of the burst on 2005 May 04, modulated by 3-min oscillations (see Fig.~\ref{fig9}). The loop-like region of the maximum emission in the spike-like pulse is also apparently located perpendicular to the coronal magnetic fan structure originating in the sunspot, which is highlighted by the 3-min narrowband map of the event on 2005 May 4. In this case, again, the energy release can be triggered by the 3-min oscillations guided by a magnetic plasma structure originating in the sunspot.

\section{Discussion}

The microwave light curves of solar flares on 2005 April 28 and 2005 May 4 contain pronounced variations with periods of about 3 min. This behaviour indicates that there is an apparent relationship with 3-min oscillations in the sunspot situated close to the flare sites.  The aim of this paper was to understand this relationship. Our analysis of dynamical features in the microwave, EUV, white light, and X-ray imaging data of AR~10756 acquired during its passage through the solar disk from 2005 April 28 to 2005 May 4, inferred the dynamical morphology of the active region. The 3-min narrowband signals detected over the sunspot and in the flare site are all well localised, which excludes their possible link with the instrumental artifacts, such as sidelobes of the image synthesis, and hence are natural. The 3-min narrowband maps of the active region, constructed with the use of PWF show the presence of extended V-shaped sources situated over the sunspot, with arms extended towards the flare site. We interpret these arms as evidence of the magnetic plasma channels that link the sunspot and the flare site by guiding magnetohydrodynamic waves. The 3-min periodicities of energy releases are then triggered by the 3-min oscillations leaking out from the sunspot along the magnetic structures.

On the basis of our findings, we deduce that the physical mechanism responsible for the relationship between 3-min sunspot oscillations and 3-min QPP in nearby flares can be as follows. The energy of 3-min oscillations leaks out of the sunspots in the form of field-aligned slow magnetoacoustic waves, which are often seen as compressible variations in the EUV radiation in the magnetic fan structures over sunspots (e.g., De Moortel 2006). In our study, these waves are seen as a 3-min modulation of the microwave radiation. The spatial distribution of the emission highlights the waveguiding plasma structures, which are the V-shaped microwave sources found in both flares discussed above. Because of the curvature of the magnetic field lines, the centrifugal force associated with the periodic longitudinal field-aligned wave motions produce periodic transverse kink-like perturbations of the magnetic structures (Zaitsev \& Stepanov 1989). The induced transverse motions are fast magnetoacoustic waves that can carry energy and information across the magnetic field. These kink waves can trigger flaring energy releases (e.g., by the mechanism proposed by Nakariakov et al. 2006) provided that the waveguiding channel is situated close to the magnetic null-point, while is not necessarily linked magnetically. The modulation depth of the flaring light curves can be significantly stronger than in the modulating signal. Hence, the leaking 3-min wave can either trigger the energy release in a form of a short aperiodic spike or spike-like pulse, as seen on 2005 April 30, or lead to periodic triggering (or modulation) of energy releases in longer duration bursts, as seen on 2005 April 28 and 2005 May 4. In the first case, the triggered energy release uses up the energy stored in the magnetic configuration, and the next maximum in the wave cannot trigger another energy release. In the second case, either not all stored energy is liberated in the previous releases, or the next period of the triggering wave causes the energy release at another spatial location (see the discussion in Nakariakov et al. 2006).

In both analysed bursts, there is observational evidence that before the flares the energy of 3-min oscillations in the sunspot is enhanced significantly. In both cases, the amplitude of 3-min wave trains was highest just before the onset of the burst. This provides interesting perspectives on the use of the increase in the power of 3-min oscillations just before the flare as a flare precursor. However, the relationship between the amplitude of 3-min oscillations in the sunspot and energy releases nearby, requires statistical proof, and should be subject to a dedicated study. 

A possible interpretation of the preferential appearance of the spike-like events when the analysed active region is in the vicinity of the central meridian, mentioned in Sect.~\ref{sec:obs}, can be interpreted as either just the time coincidence with the appearance of certain physical conditions for the energy releases (e.g., the emergence of a new loop or arcade of loops from the sunspot to the burst source) or the preferential observability conditions (e.g., connected with the line-of-sight angle). The latter issue could be indicative of the similarity between the observed pulses and well-known microwave spikes: Altyntsev et al. (1996) found out that the spatial size of the microwave spike source is systematically larger in the vicinity of the limb. This was interpreted in terms of the scattering of the microwave emission across the coronal plasma. Hence, some events situated close the limb can be of lower intensity due to the scattering and hence be below the detection threshold and missed by observations. This may explain the preferential appearance of the short pulses in the vicinity of the central meridian.

\begin{acknowledgements}
A part of this work was supported by the Royal Society UK-Russian International Joint Project, the Grant No. 300030701
of Grant Agency of the Czech Academy of Sciences, and the grants RFBR 08-02-13633-ofi-c, 08-02-91860-KO-a, 08-02-92204-GFEN-a.
\end{acknowledgements}



\end{document}